\documentclass[%
 amsmath,amssymb,amsfonts,
 aps,prl,twocolumn
]{revtex4-1}

\usepackage{amsmath}
\usepackage[dvips]{graphicx}% Include figure files
\usepackage{wrapfig}
\usepackage{bm}% bold math
\usepackage{overpic}

\newcommand{\FDir}{./}
\newcommand{\scs}{\scriptsize}

\newcommand{\pll}{\parallel}
\newcommand{\pp}{\perp}
\newcommand{\nn}{\nonumber}

\begin{document}

%\preprint{APS/123-QED}

\title{First Principles Study of the Optical Dipole Trap for Two-Dimensional Excitons in Graphane}% Force line breaks with \\
\thanks{A footnote to the article title}%

\author{$^1$Hiroki Katow, $^2$Ryosuke Akashi, $^3$Yoshiyuki Miyamoto, and $^2$Shinji Tsuneyuki }
\affiliation{$^1$Photon Science Center, Graduate School of Engineering, The University of Tokyo, Bunkyo-ku, Tokyo 113-8656, Japan}
\affiliation{
$^2$Department of Physics, The University of Tokyo, 7-3-1 Hongo, Bunkyo-ku, Tokyo 113-8656, Japan
}%
\affiliation{%
$^3$Research Center for Computational Design of Advanced Functional Materials, National Institute
of Advanced Industrial Science and Technology (AIST), Central 2, Tsukuba, Ibaraki 305-8568, Japan
}%
%\affiliation{
%\affiliation{%

\date{\today}% It is always \today, today,

\begin{abstract}
%Recent developments of the research for excitons in two-dimensional materials are strongly driven by their potential for novel electronic and optical devices. 
Recent studies on excitons in two-dimensional materials have been widely conducted for their potential usages for novel electronic and optical devices. 
Especially, sophisticated manipulation techniques of quantum degrees of freedom of excitons are demanded.
In this paper we propose a technique of forming an optical dipole trap for excitons in graphane, a two-dimensional wide gap semiconductor, based on first principles calculations.
We develop a first principles method to evaluate the exciton transition dipole matrix and combine it with the density functional theory and GW+BSE calculations. 
We reveal that in graphane the huge exciton binding energy and the large dipole moments of Wannier-like excitons enable us to induce the dipole trap of the order of meV depth and $\mu$m width.
This work opens a new way to control light-exciton interacting systems based on a newly developed numerically robust {\it ab initio} calculations. 
%gauge invariant な light-exciton interacting systemの第一原理からのモデリングを
%We developed a method for calculating the exciton transition dipole matrix which is independent from the choice of gauge of the Kohn-Sham orbitals. 
\end{abstract}

\maketitle

%\tableofcontents
%introduction
%exciton background
% 2D system 
% rise of TMD
The exciton is a quasi particle formed by a pair of electron and hole in solids.
The binding and relaxation processes of the exciton dominate the optical response of the semiconductors. 
The Bose-Einstein condensate of excitons has also been of much interest as a novel quantum many-body phase in virtue of its extremely lighter mass than nuclei.
Optical control of excitons via the strong light-exciton coupling is particularly an intriguing subject for its application to optical devices. 
For example, through the studies on transition metal dichalcogenides (TMD), the valley-selective formation of the excitons by circularly polarized light and their transport have been intensively pursued \cite{Review16, Review18}. 
Further sophistication of the exciton manipulating techniques  will benefit the future development of the field.

The subject of this study is a spatial control of the center-of-mass motion of the excitons. Let us list a few pioneering studies on this:
applying a uniaxial strain potential by mechanical force \cite{BHSPW07-Strain, YCK11-BECStrain},
stacking a periodic metallic gates on a quantum well \cite{Letal07-MetalGate}, %30
utilizing a repulsive potential between spatially indirect excitons \cite{APAHGCB10-QW,Wetal10-QW,HGBSIG06-QW}, %31-33
inducing a periodic strain field by surface acoustic wave (SAW) \cite{SKGVLC17-SAW, LPSH06-SAW,IL01-SAW}, %34-36
utilizing the Pauli repulsion from virtually excited excitons \cite{LB03-Pauli, BL06-Pauli,CMP11-Pauli,CMP11-Pauli}, %37-40
and making use of the optical Stark shift of trion resonance for free career trapping \cite{CMP11-ExOpt}.%41

Here, we consider application of the idea in the research field of ultracold atomic systems. Rich variety of electromagnetic techniques have been implemented for designing the geometry of the confinement potential, which have been already applied to the studies on quantum many-body phases like the atomic BEC and the quantum simulator \cite{GZ14,GZ15}. 
A representative confinement mechanism in practice is the energy level shift induced by electromagnetic fields whose frequencies are nearly resonant with the electronic energy gap of atoms; this is called optical dipole trap. 
Our central object is the optical dipole trapping phenomenon in the excitonic system, which could provide us with a precise non-mechanical control of the excitons like that of the cold atomic systems.

In this paper we theoretically propose an optical technique to control the center-of-mass motion of free excitons in graphane (hydrogen-terminated graphene), a two-dimensional wide gap semiconductor.
The optical potential for the excitons in graphane can be implemented as a spatial energy shift of the exciton dressed state which is formed by irradiating external laser fields.
We show that two- and three-level systems are well defined in this system with the use of first principles methods, namely the density functional theory (DFT), GW approximation and the Bethe-Salpeter equation.
The dressed states are parametrized by the electronic transition dipole moment and energy level configuration of graphane excitons.
The dipole parameter in the Bloch representation requires derivatives in the reciprocal space, which generally suffers from the gauge dependence. We constructed a gauge-free formula which is analogous in spirit to the King-Smith--Vanderbilt formula for  electronic polarizability calculations in ground states\cite{KV93,R94}.
We found that some features of graphane excitons -- huge binding energy, wide energy level spacing, and large transition dipole moment -- yields the deep potential shift of order of meV without violating the rotational wave approximation \cite{GZ14,GZ15}.
These features are common in two-dimensional semiconductors such as TMDs. Thus we can expect a wide application of this technique.

We begin with a general introduction of laser-matter interaction in a few-level system with the dipole approximation.
In particular we consider a specific three-state system where two out of three levels are degenerate, which is later seen to correspond to graphane.
Applying the rotational wave approximation, the system is described by the Rabi Hamiltonian $H_{R}$.
Eigenstates of $H_R$ are called dressed states.
The Rabi Hamiltonian of a two-level system $H_{R}^{(2)}$ and its eigenvalues $E^{\pm}$ are given by
\begin{eqnarray}
H^{(2)}_R &=& 
\begin{pmatrix}
-\Delta & -\Omega_R/2 \\
-\Omega_R^*/2 & 0
\end{pmatrix}\\
E_{\pm} &=& -1/2 \{\Delta \pm (\Delta^2 + |\Omega_R|^2)^{1/2}\},
\end{eqnarray}
in atomic unit.
Here $\Delta$ is detuning and $\Omega_R=2\bm{E}\cdot \bm{d}_{12}$ is the Rabi frequency. 
$\bm{E}$ is a local external field and $\bm{d}_{12}$ is electric transition dipole moment which is defined by a matrix $\bm{d}_{S'S} = \langle S'|\bm{r}|S\rangle $ where $|S(S')\rangle$ is the excitonic eigenstate.
In the same way the Rabi Hamiltonian of a three-level system $H^{(3)}_R$ and corresponding eigenvalues are derived as follows:
\begin{eqnarray}
H^{(3)}_R &=& 
\begin{pmatrix}
0 &0 &  \Omega_{13}/2 \\
0 & 0 & \Omega_{23}/2 \\
 \Omega^*_{13}/2 & \Omega_{23}^*/2 & -\Delta
\end{pmatrix}
\\
E_{0,\pm} &=& 0, -1/2 \{\Delta \pm (\Delta^2 + |\Omega_{13}|^2 + |\Omega_{23}|^{2})^{1/2}\}
\end{eqnarray}
Here $\Omega_{13(23)}= 2\bm{E}\cdot \bm{d}_{13(23)}$ is the Rabi frequency.
We assumed that among three excitonic states $|S\rangle$ ($S$=1,2,3) the first two are degenerate.
Thus in both cases the system is parametrized by $\bm{E}$, $\Delta$, and $\bm{d}$.
The transition dipole moment $\bm{d}$, being the only intrinsic quantity, must be given by {\it ab initio} calculations.
We referred to \cite{GZ14, GZ15} for above formulations.
$U_{op}^{\pm}= E_{\pm}(\bm{E})-E_{\pm}(\bm{0})$ gives optical potential depth which moderately varies in the spatial scale of laser wave length as $U_{op}^{\pm}(\bm{r})$.

Calculation of the dipole moment in a periodic system needs a special care as there is an ambiguity in definition of the unit cell boundary and  the dependence on a gauge of the Kohn-Sham orbital.
King-smith and Vanderbilt have reformulated the Berry phase formula for the electric polarization of dielectrics which is free from the choice of gauge \cite{KV93,R94}.
A similar problem is present in the transition dipole moment of the excitonic eigenstates in the periodic system.
In this work we extend the King-Smith-Vanderbilt formula for excitonic systems as follows:
\begin{eqnarray}
d\bm{k} \cdot \langle S' |\bm{r}|S\rangle &\simeq& \sum_{\bm{k}_{\pp}}\Im [\ln \prod_{\bm{k}_\pll}\{ \det \{T^{SS'}_{e^-}+W_{e^-}\}\nn \\
&-&\ln \prod_{\bm{k}_\pll}\{ \det \{T^{SS'}_{ h^+}+W_{ h^+}\}].
\label{eq:BP}
\end{eqnarray}
Here $|S\rangle = \sum A^s_{kvc}a^\dagger_{kc} a_{kv}|{\rm g.s.}\rangle$ is an excitonic state, 
\begin{eqnarray}
T^{S'S}_{e^-,c_1c_2} &=& \sum_{c_3}^{N_c}\sum_{v}^{N_v}A^{S'*}_{\bm{k},vc_3}A^{S}_{\bm{k},vc_1}\langle u_{\bm{k},c_3} |u_{\bm{k}+d\bm{k},c_2}\rangle\\
T^{S'S}_{h^+,v_1v_2} &=& \sum_{c}^{N_c}\sum_{v_3}^{N_v}A^{S'*}_{\bm{k},v_1c}A^{S}_{\bm{k},v_3c}\langle u_{\bm{k},v_3} |u_{\bm{k}+d\bm{k},v_2}\rangle\\
W_{e^-(h^+)mn} &=& \delta_{mn}\langle u_{\bm{k},m}| u_{\bm{k}+d\bm{k},m}\rangle/| \langle u_{\bm{k},m}| u_{\bm{k}+d\bm{k},m}\rangle|.
\end{eqnarray}
$|u_{\bm{k},m}\rangle$ is a cell periodic factor of the Kohn-Sham orbital.
$N_c$ is the number of conduction bands and $N_v$ is the number of valence bands.
The product $\prod_{\bm{k}_{\pll}}$ is taken along a closed path parallel to $d\bm{k}$ in BZ and $\sum_{\bm{k}_{\pp}}$ sums up all the paths.
$d\bm{k}$ is a vector whose norm is equal to the $k$ point spacing and is parallel to reciprocal lattice vectors in our calculation.
Index of the matrices with a subscricpt $e^-$ ($h^+$) runs over conduction (valence) bands.  
Eq.~(\ref{eq:BP}) still possesses a multivalued nature of logarithm given by an integer multiple of $2\pi$.
Its contribution can be omitted as long as we sample large number of $k$ points. 
See \cite{Supp} for details.

\begin{figure}
\begin{tabular}{ccc}
	\begin{minipage}{0.25\hsize}
	\begin{center}
	\begin{overpic}[angle=90,width=2cm]{\FDir/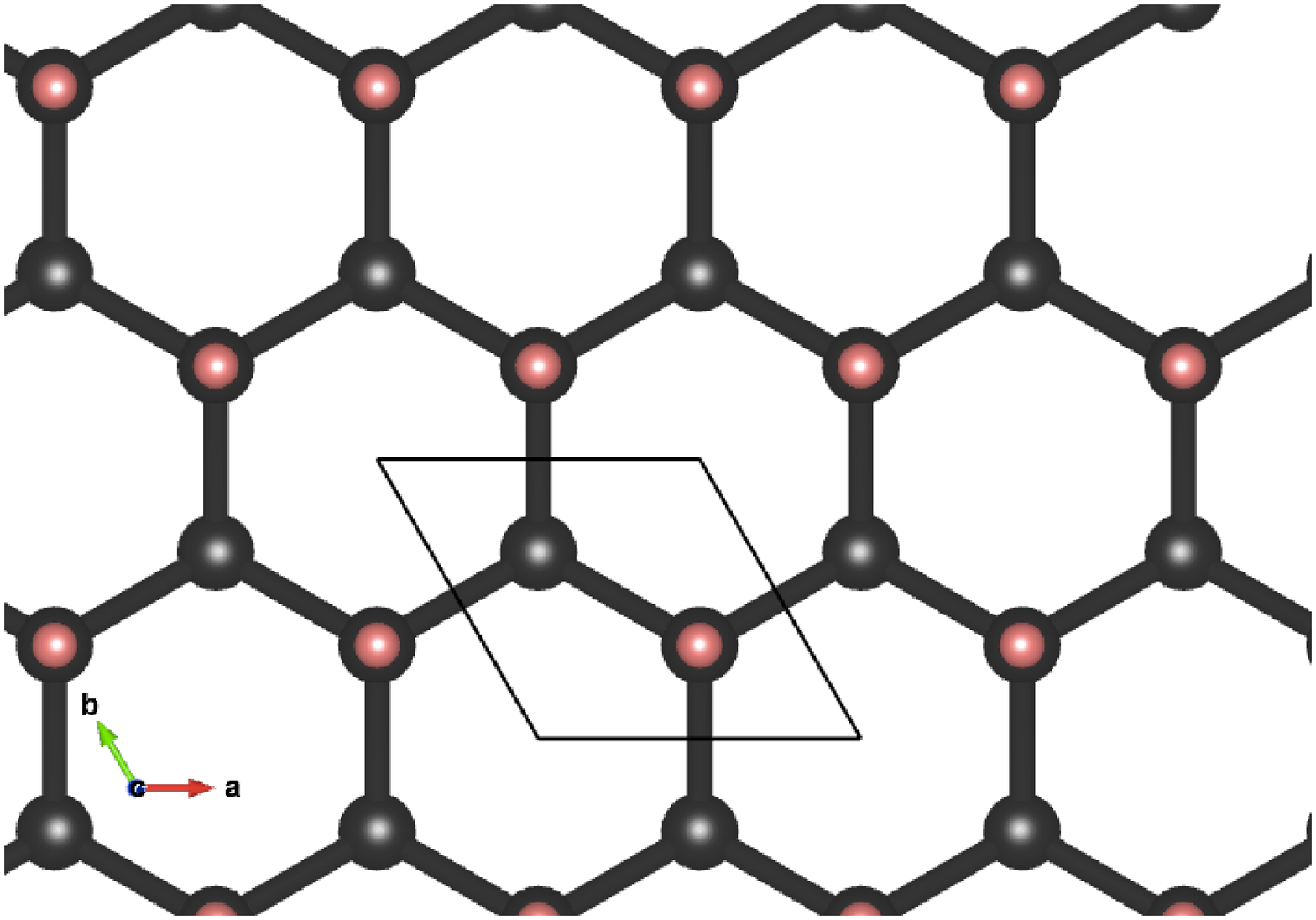}
	\put(0,73){\fbox{\includegraphics[clip, width=1.0cm]{\FDir/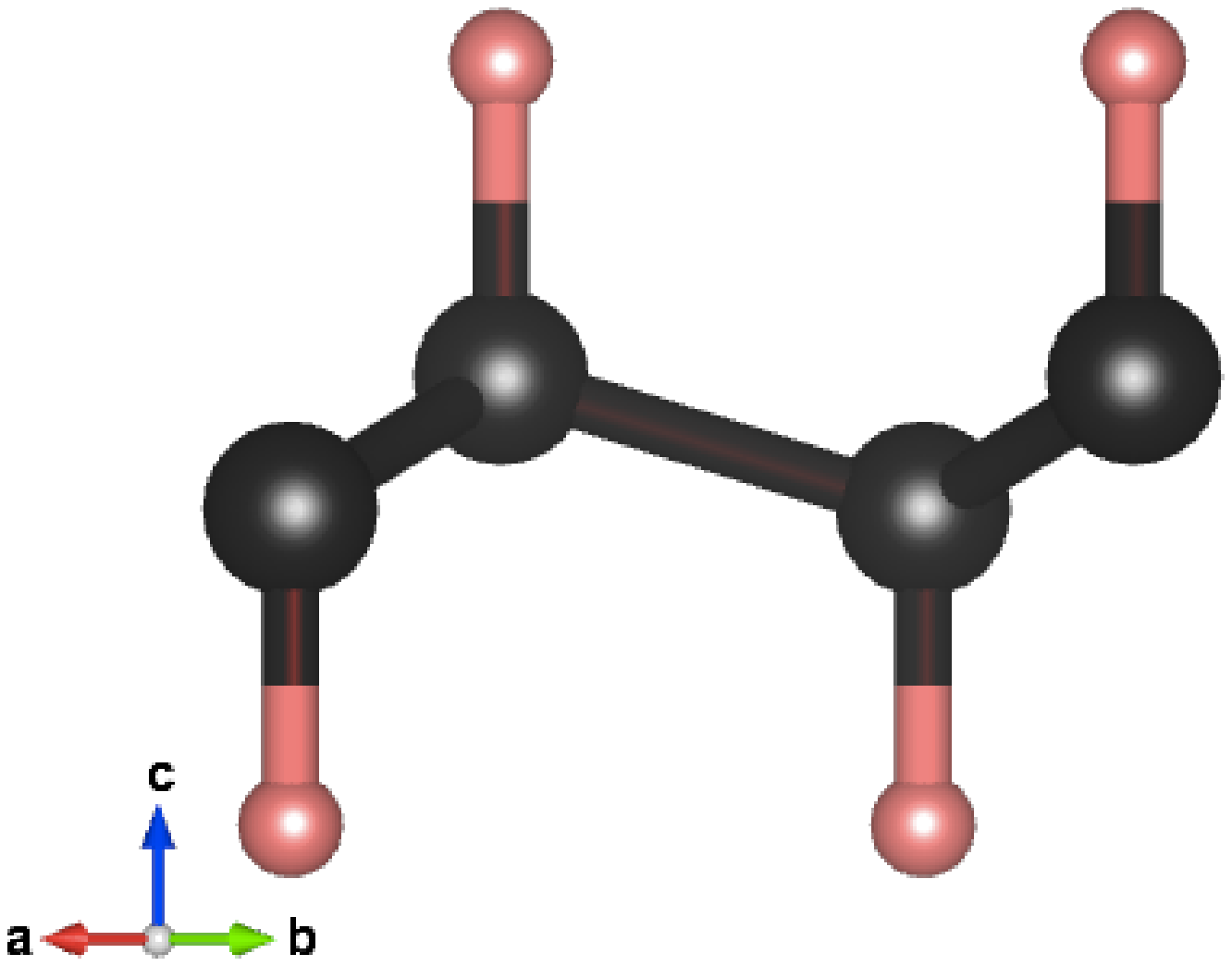}}}	
	\put(-15,88){(a)}
	\end{overpic}
	\end{center}
	\vspace{-1mm}
	\end{minipage}
	&
	\begin{minipage}{0.5\hsize}
	\begin{center}
	\begin{overpic}[percent,width=5cm]{\FDir/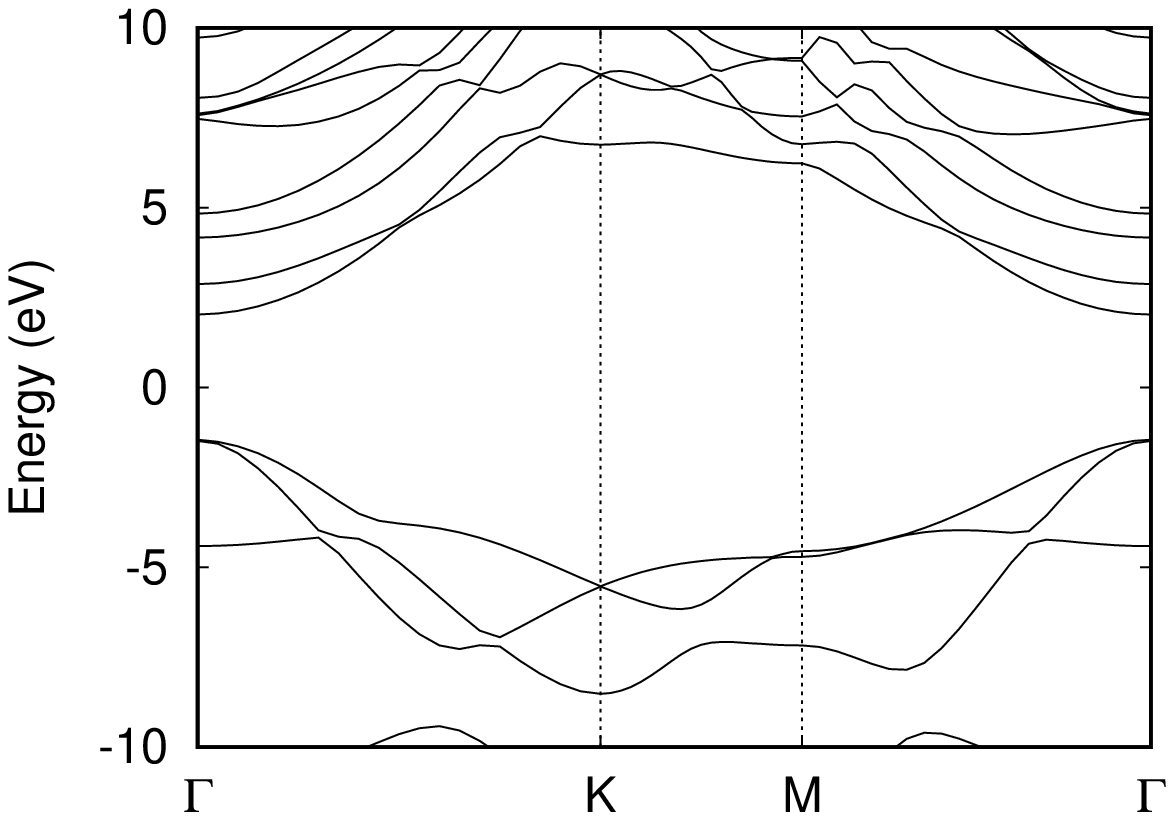}
		\put(0,60){(b)}
	\end{overpic}
	\end{center}
	\end{minipage}
\end{tabular}
\caption{(a) The crystal structure of the chair-type graphane seen from (001) direction and (110) direction (inset). The black (pink) balls indicate carbon (hydrogen) atoms. A wedge-shaped area indicates a unit cell. (b) The LDA band structure of the graphane calculated by using the local density approximation. 
} 
\label{fig:structure}
\end{figure}

Graphane is a two-dimensional wide gap semiconductor composed of hydrogenated graphene. 
Amongst possible metastable states, the \text{\it chair-type} conformation is reported to be most stable \cite{SLSP15}.
We show the crystal and DFT band structure of \text{\it chair-type} graphane in FIG.~\ref{fig:structure}(a), (b).
Graphane possesses $D_{3d}$ group symmetry and hence has no permanent dipole. 
Accordingly the degree of the degeneracy can not be more than double, which greatly simplifies further analysis.
A direct Kohn-Sham gap opens at $\Gamma$ point whose width is 3.4 eV.
DFT calculation was done by {\it Quantum-ESPRESSO} package\cite{G09} with 18$\times$18$\times$1 $k$ points and 150 Ry energy cutoff.
PBE type energy density functional was used \cite{GTH1996, HGH1998}. 
It is well known that the DFT calculation systematically underestimates the energy gap.
We therefore  implemented energy level calculations based on GW approximation which takes the screening effect by medium into account.
The GW calculation and subsequent BSE calculation were done by using {\it BerkeleyGW} code \cite{DSSJCL12}.
Resulting energy band gap is 6.8 eV.
Detailed calculation conditions and the band gap convergence is summarized in supplement \cite{Supp} in comparison with preceding works \cite{CATR10,LKEK09,KZO12,LPHPP10}.

\begin{figure}
	\begin{tabular}{lc}	
		\begin{minipage}{0.45\hsize}
		\begin{center}
		\begin{overpic}[width=4 cm]{\FDir/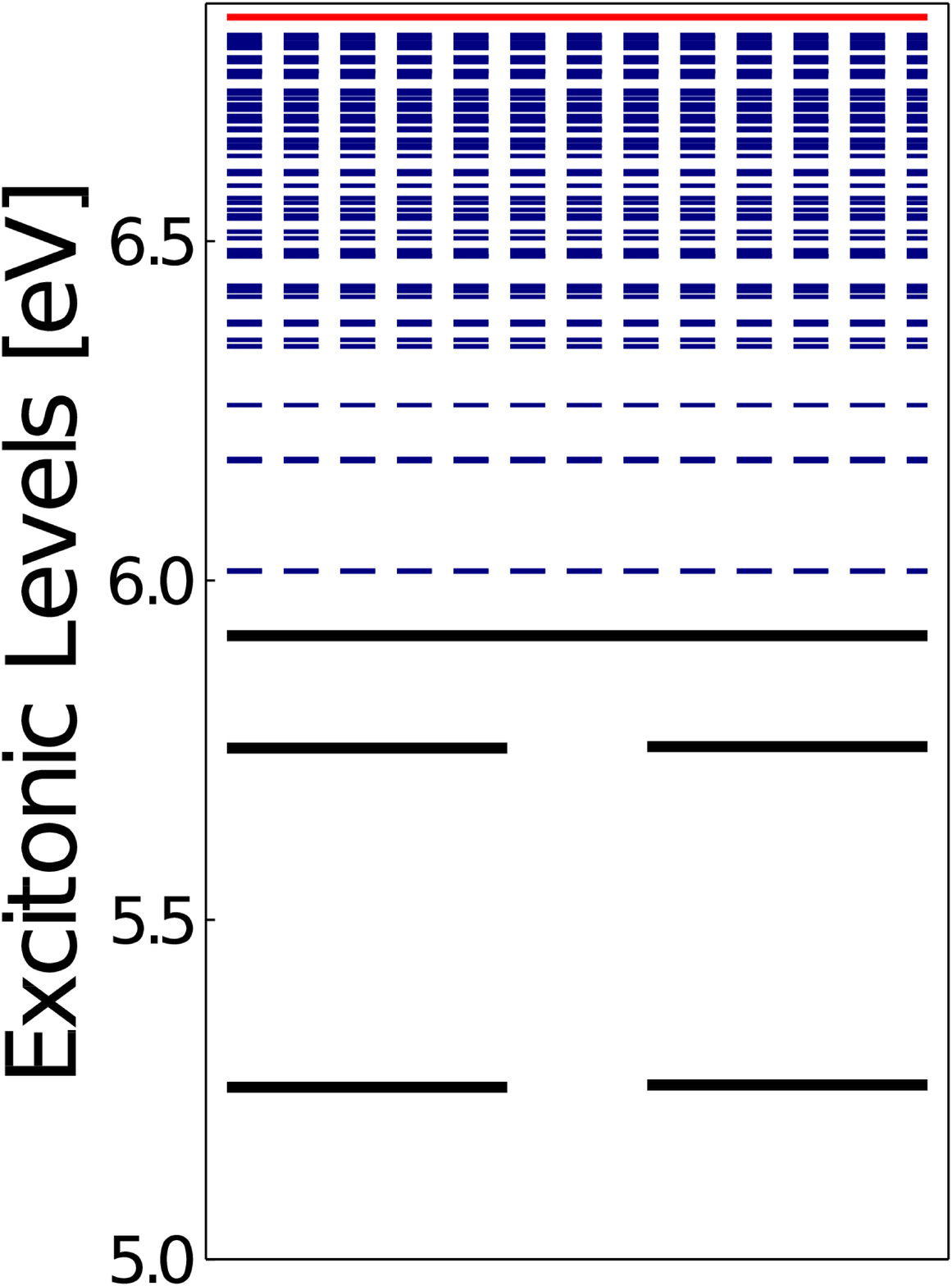}
		\put(0,90){(a)}
		\put(20,10){\scs $E_u$(5.254)}		
		\put(50,10){\scs $E_u$(5.257)}		
		\put(20,35){\scs $E_g$(5.753)}		
		\put(50,35){\scs $E_g$(5.755)}		
		\put(35,45){\scs $A_{2g}$(5.919)}	
		%\put(5,90){\scs CBM}	
		\end{overpic}
		\end{center}
		\end{minipage}
	&
		\begin{tabular}{|cc|}
		\hline
			\begin{minipage}{0.25\hsize}
			\vspace{1mm}
			\begin{center}
			\begin{overpic}[clip,width=2.2cm]{\FDir/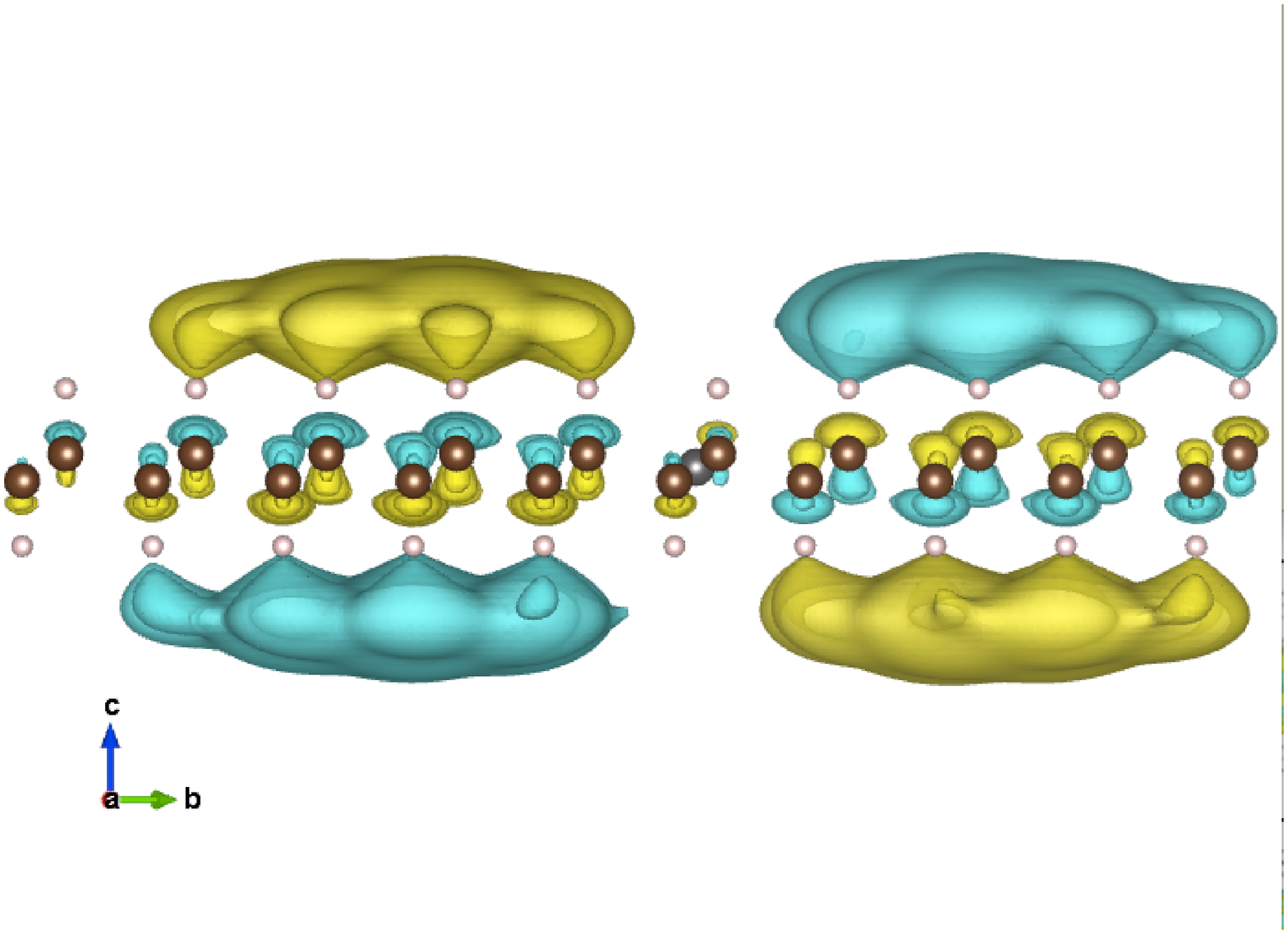}
			\put(0,63){(b)$A_{2g}$}
			\end{overpic}
			\end{center}
			\end{minipage}
		&			
			\begin{minipage}{0.25\hsize}
			\begin{center}
			\begin{overpic}[clip, width=2.2 cm]{\FDir/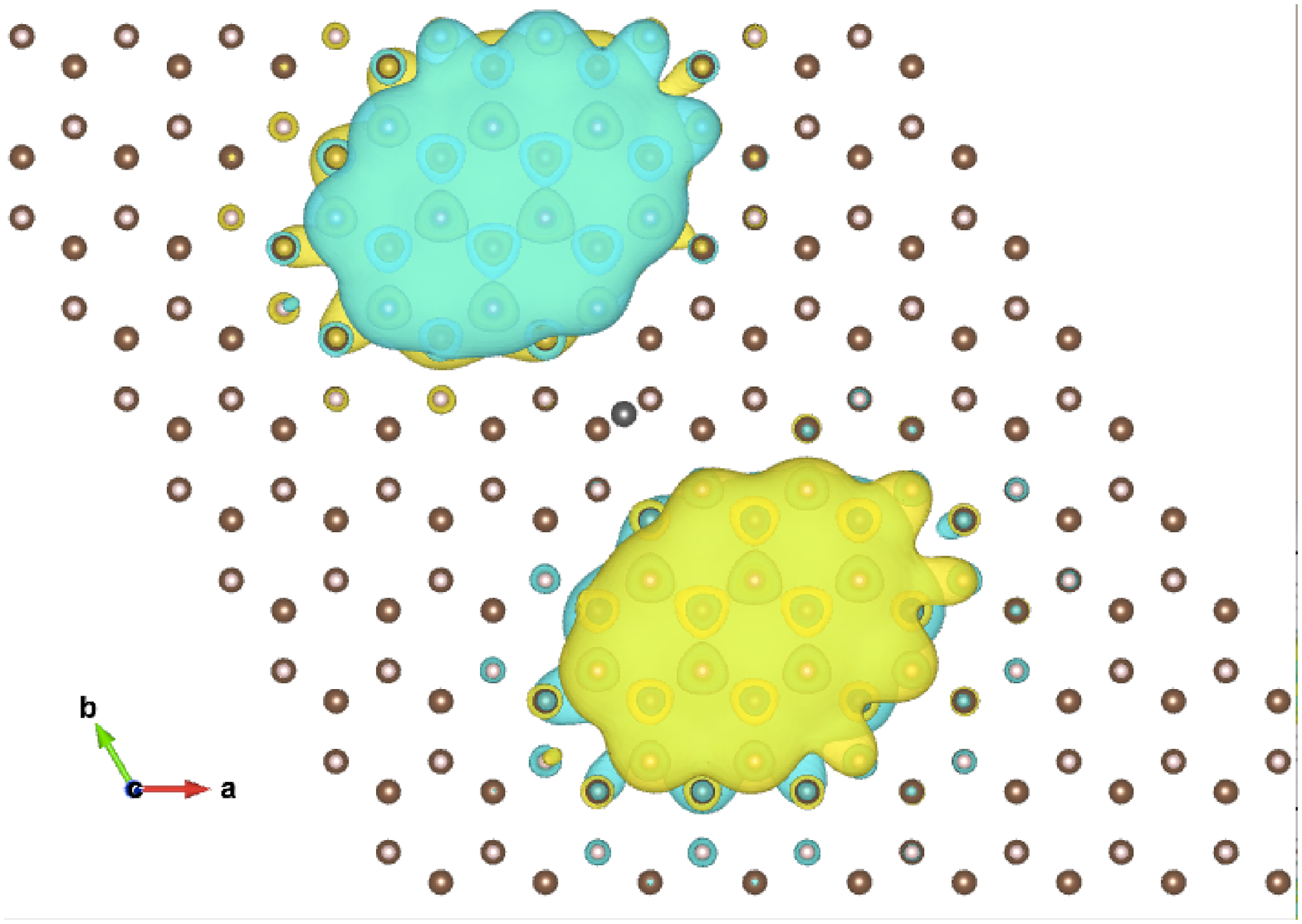}			
			\end{overpic}
			\end{center}
			\end{minipage}
		\\	
		\hline	
			\begin{minipage}{0.25\hsize}
			\vspace{1mm}
			\begin{center}
			\begin{overpic}[clip, width=2.2 cm]{\FDir/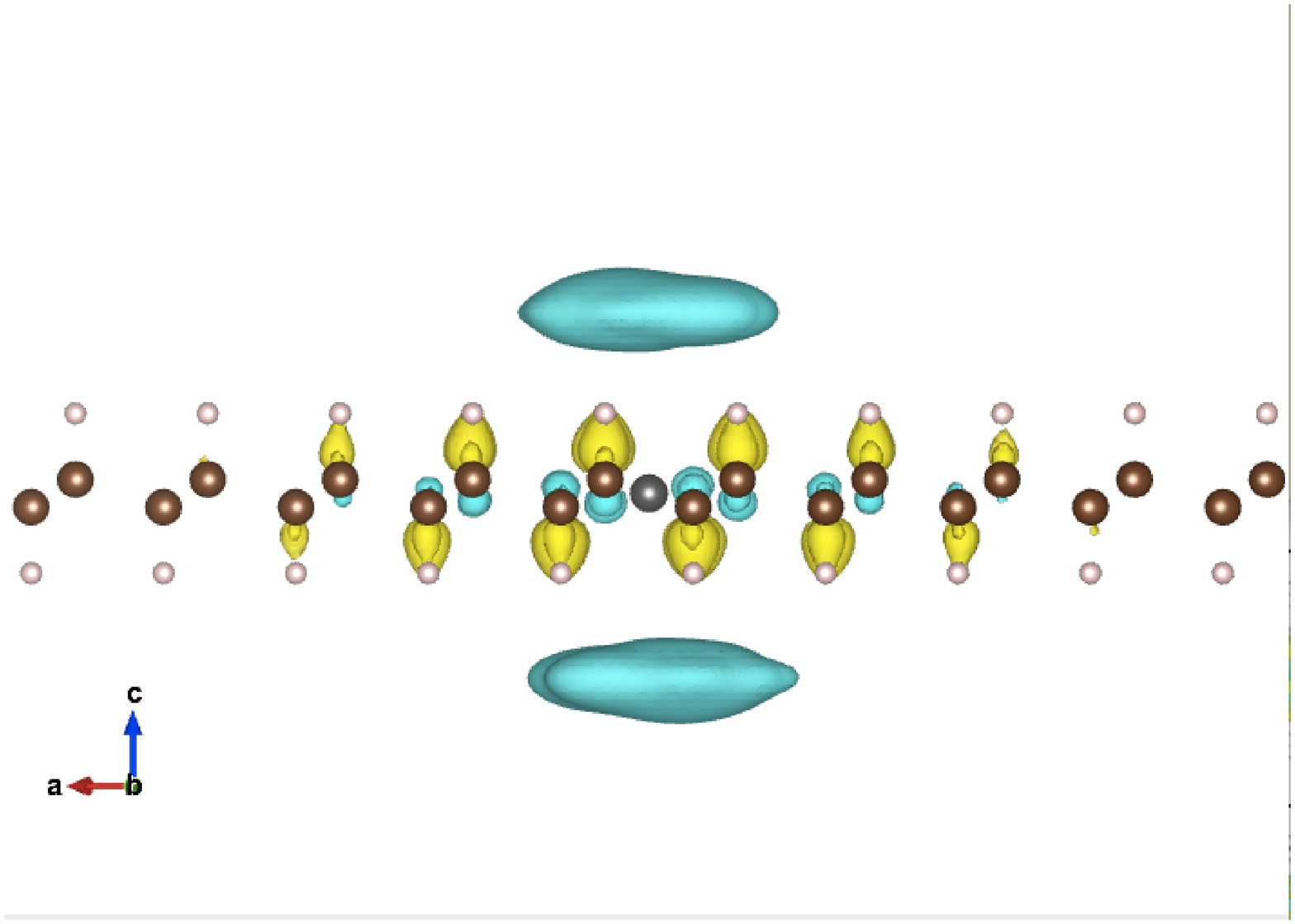}
			\put(0,63){(c)$E_{g}$}
			\end{overpic}
			\end{center}
			\end{minipage}
		&
			\begin{minipage}{0.25\hsize}
			\begin{center}
			\begin{overpic}[clip, width=2.2 cm]{\FDir/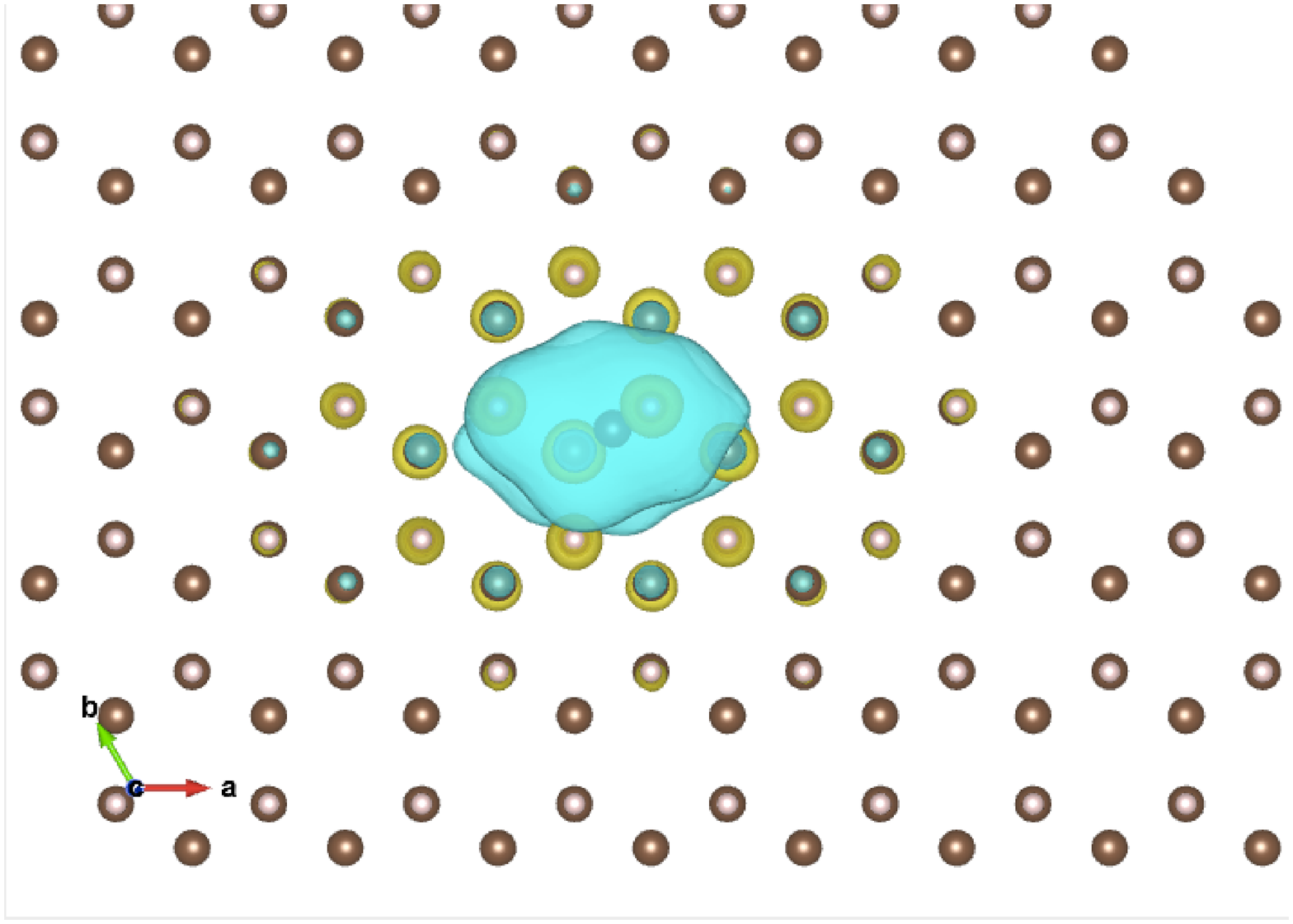}
			\end{overpic}
			\end{center}
			\end{minipage}
		\\
		\hline
			\begin{minipage}{0.25\hsize}
			\vspace{1mm}
			\begin{center}
			\begin{overpic}[clip, width=2.2 cm]{\FDir/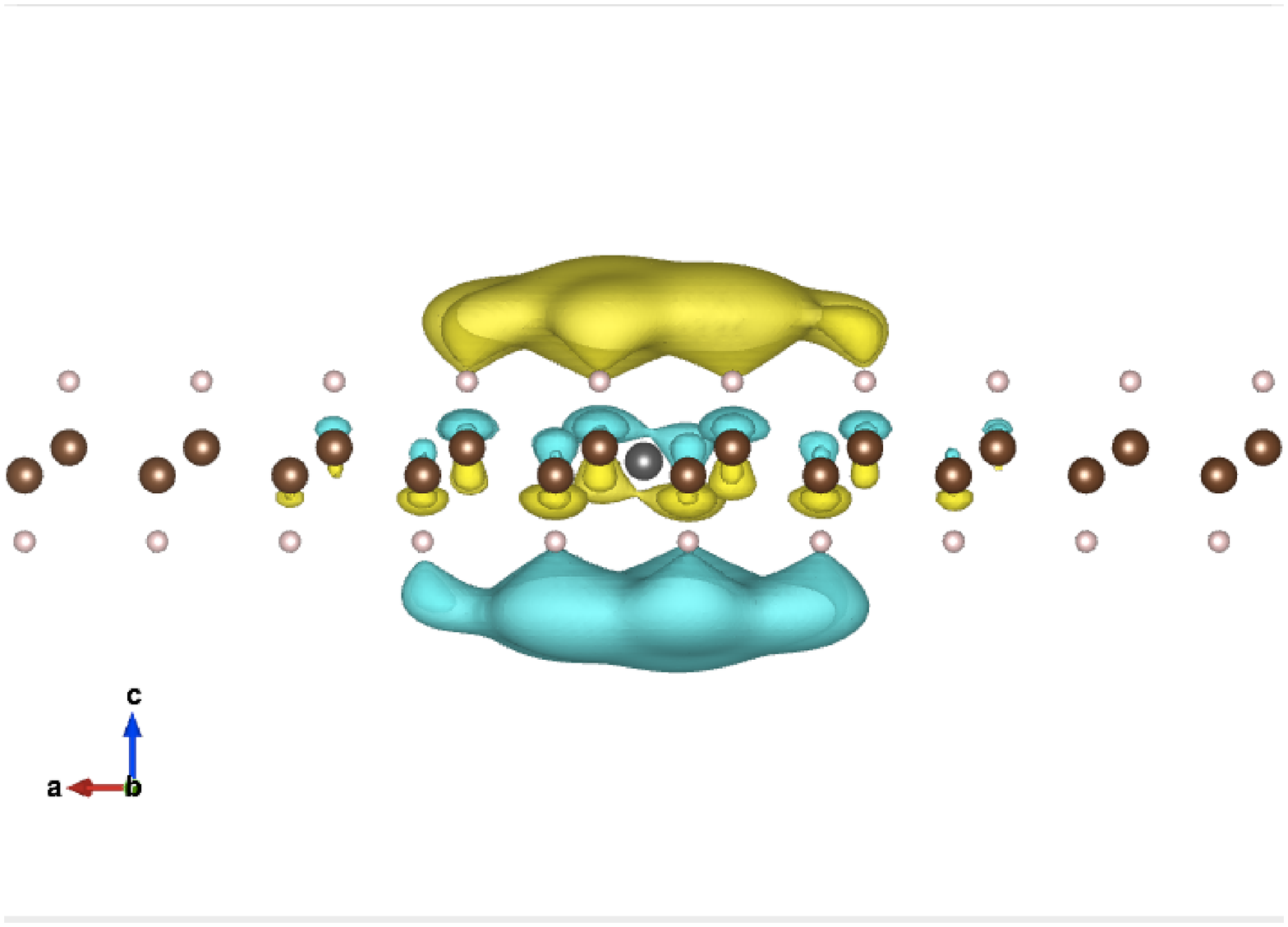}
			\put(0,63){(d)$E_{u}$}
			\end{overpic}
			\end{center}
			\end{minipage}
		&
			\begin{minipage}{0.25\hsize}
			\begin{center}
			\begin{overpic}[clip, width=2.2 cm]{\FDir/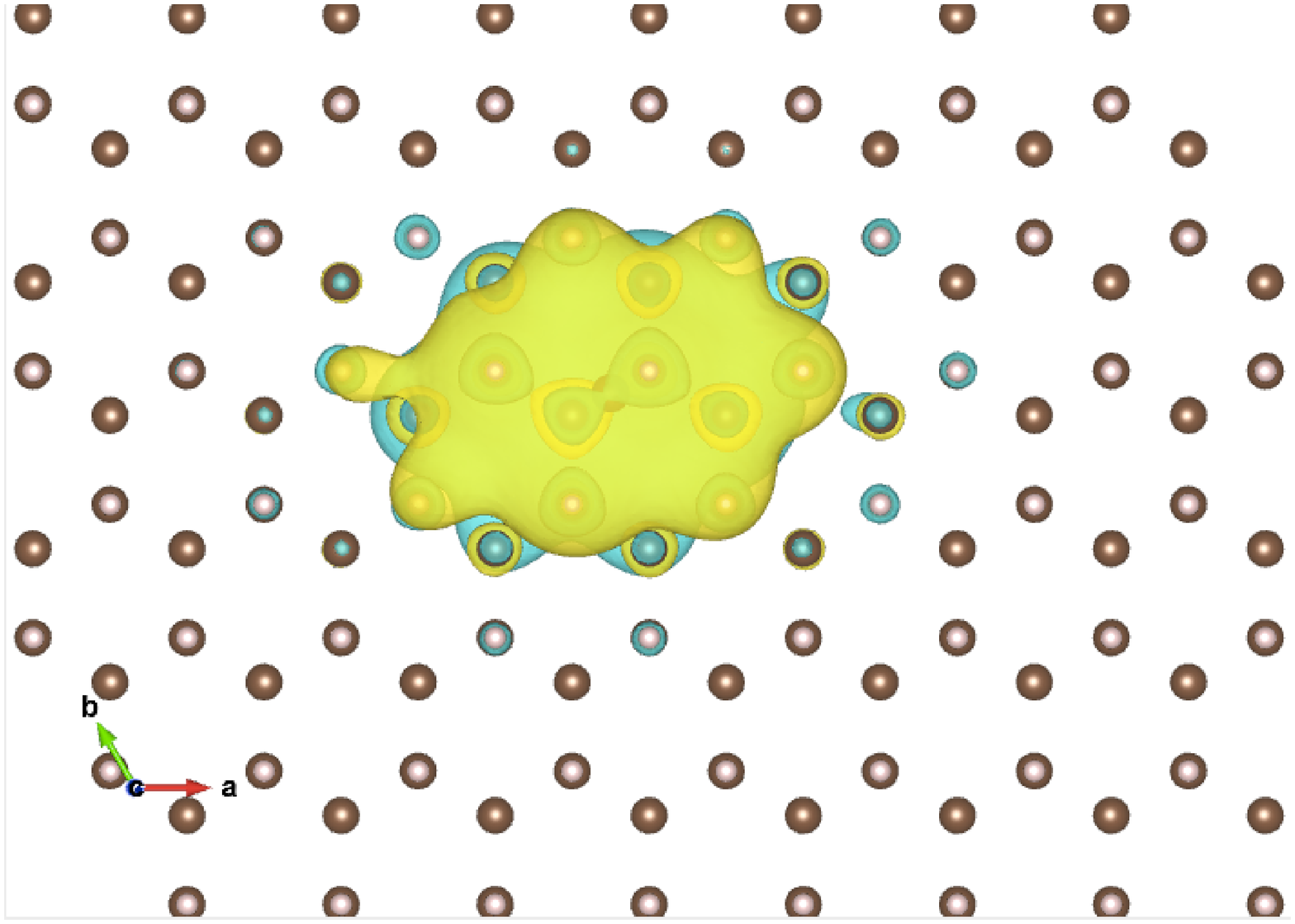}
			\end{overpic}
			\end{center}
			\end{minipage}
		\\
		\hline
		\end{tabular}
	\end{tabular}
	\caption{(a) Excitonic levels measured from the valence band maximum (VBM) in eV on the $\Gamma$ point. Characters of the $D_{3d}$ group which each level belongs to are shown for the first five levels with its energy in brackets. The CBM is at 6.8 eV and shown by a red solid line. Real part of the excitonic wave function is depicted for (b) the $A_{2g}$ exciton, (c) $E_g$ exciton, and (d) the $E_u$ exciton from (100) (left panel) and (001) (right panel) directions. Hole coordinates are fixed in the middle of $C$-$C$ bonding. Two colors of isosurface indicate the sign of wave functions. }
	\label{fig:levels}
\end{figure}

A typical feature of two dimensional semiconductors is the weak screening  of Coulomb interaction by medium which leads to the formation of excitons with huge binding energy. 
In FIG.~\ref{fig:levels}(a) energy levels calculated by the GW+BSE method with the Tamm-Dancoff approximation
 \footnote{A. Fetter and J. D. Walecka, {\it Quantum Theory of Many Particle Systems} (McGraw-Hill Book Company, San Francisco, 1971), pp. 538-539.} are plotted. 
We used 36$\times$36$\times$1 $k$ point mesh.
Lowest five levels are highlighted by bold black lines as we focus on the manipulation of low energy levels.
These five levels belong to character $E_u$, $E_g$, and $A_{2g}$ of $D_{3d}$ group.
$E_u$ and $E_g$ states are plotted by short lines to emphasize the double degeneracy.
$E_u$ exciton possesses huge binding energy which amounts to 1.6 eV measured from conduction band minimum (CBM) given by the GW calculation. 
In FIG.~{\ref{fig:levels}}(b), (c), and (d), real part of the excitonic wave functions for the lowest five levels are shown. 
The $E_u$ exciton in this work obviously corresponds to $A$ and $B$ excitons in \cite{CATR10}.
See also
\footnote[1]{A tiny gap breaking the degeneracy of $E_u$ excitons is reported in \cite{CATR10} as well, we however point out that this is probably due to the symmetry violation by $k$ point interpolation scheme in \cite{DSSJCL12}. We confirmed the gap vanishes without such interpolation in 18$\times$18$\times$1 $k$ point calculation. We hence neglect the gap in this paper.}.

\begin{table}[htb]
\begin{center}
	\begin{tabular}{c}
	
	\begin{minipage}{1.0\hsize}
	\begin{tabular}{ccccc}
		S & S' & $\sqrt{{\rm tr} D_{xx}}$ & $\sqrt{{\rm tr} D_{yy}}$ & $\sqrt{{\rm tr} D_{zz}}$\\ \hline \hline
		$E_u$ & $E_g$ & 0.13 & 0.18 & 4.59 \\
		$E_u$ & $A_{2g}$ & 7.87 & 13.3 & 0.00 \\
		$E_u$ & $E_g^{(2)}$ & 15.1 & 20.4 & 0.00 \\
	\end{tabular}	
	\end{minipage}
	
	\end{tabular}
	\caption{
	The traces of $D_{ij} = d^{i\dagger}d^j$, (i, j = $x, y, z$), where $D_{ij}$ is $2\times2$ or $2\times1$ matrix in our case. ${\rm tr }D_{ij}$ is invariant under a unitary transformation of excitonic states. }  
	\label{tab:dipole}
	\end{center}
\end{table}

The resulting transition dipole moment $\bm{d}_{SS'}$ for $S=E_u$ and $S'=E_g, A_{2g}$ are summarized in TABLE~\ref{tab:dipole}.
$S' = E^{(2)}_g$ case, where $E^{(2)}_g$ is the second $E_g$ states being adjacent to $A_{2g}$ state, is also shown for later discussions. 
When the system possesses time-inversion symmetry, the dipole matrix can always be taken as real.   
Determinant and trace of a matrix $D_{ij} = d^{i\dagger}d^{j}$ ($i,j = x, y, z$)  are unitary invariant.
These values are comparable with those of excitonic system in Quantum Dots (QDs) and one order larger than atomic systems \cite{ESNML00, SLSGKPPS01}.
As for the dipole matrix of the $E_u$ and $E_g$ states, the $z$ component is dominant and in-plane component is negligible.
 As we can see the raw values of $\bm{d}$ in \cite{Supp}, dipole matrix is then block diagonal and split into two $2\times2$ matrices.  
This block diagonalization can always be done by a unitary transformation, and thus we can construct two two-level systems from the $E_u$ and $E_g$ excitons.
The $E_u$ and $A_{2g}$ excitons form a three-level system and in-plane dipole matrices dominate in contrast to the previous case.
The second $E_g$ state is present 100 meV above the $A_{2g}$ state and it provides an upper limit of the applicable field strength  within the rotational wave approximation.
In later discussions, we consider a two-level and a three-level system using combinations of the $E_u$-$E_g$ and $E_u$-$A_{2g}$ excitonic levels, respectively.

\begin{figure}
	\vspace{5mm}
	\begin{tabular}{cc}
		\begin{minipage}{0.5\hsize}
		\begin{center}
		\begin{overpic}[clip,width=4.4 cm]{\FDir/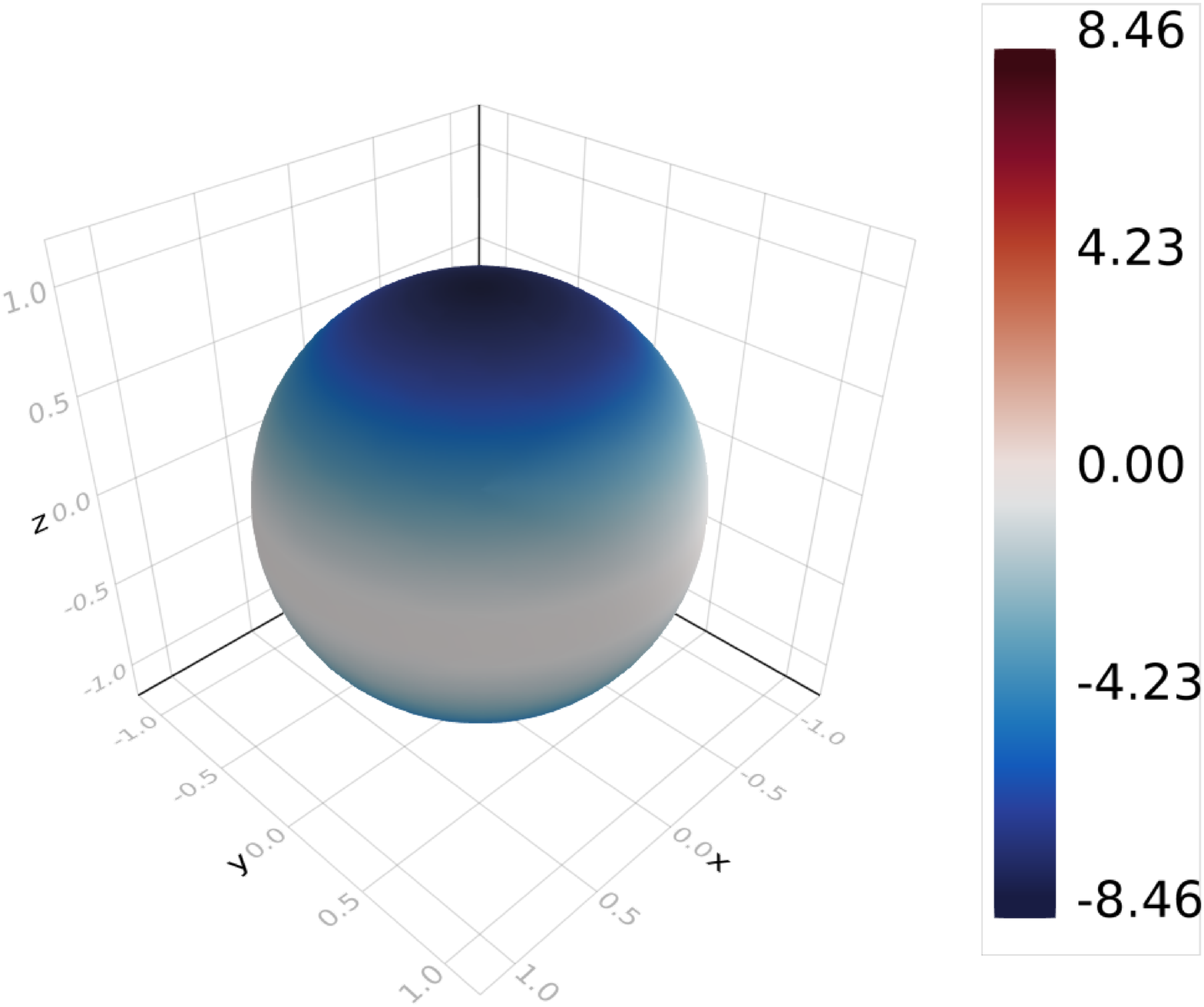}
		\put(0,73){(a)}
		\end{overpic}
		\end{center}
		\end{minipage}		
	&
		\begin{minipage}{0.5\hsize}
		\begin{center}
		\begin{overpic}[clip,width=4 cm]{\FDir/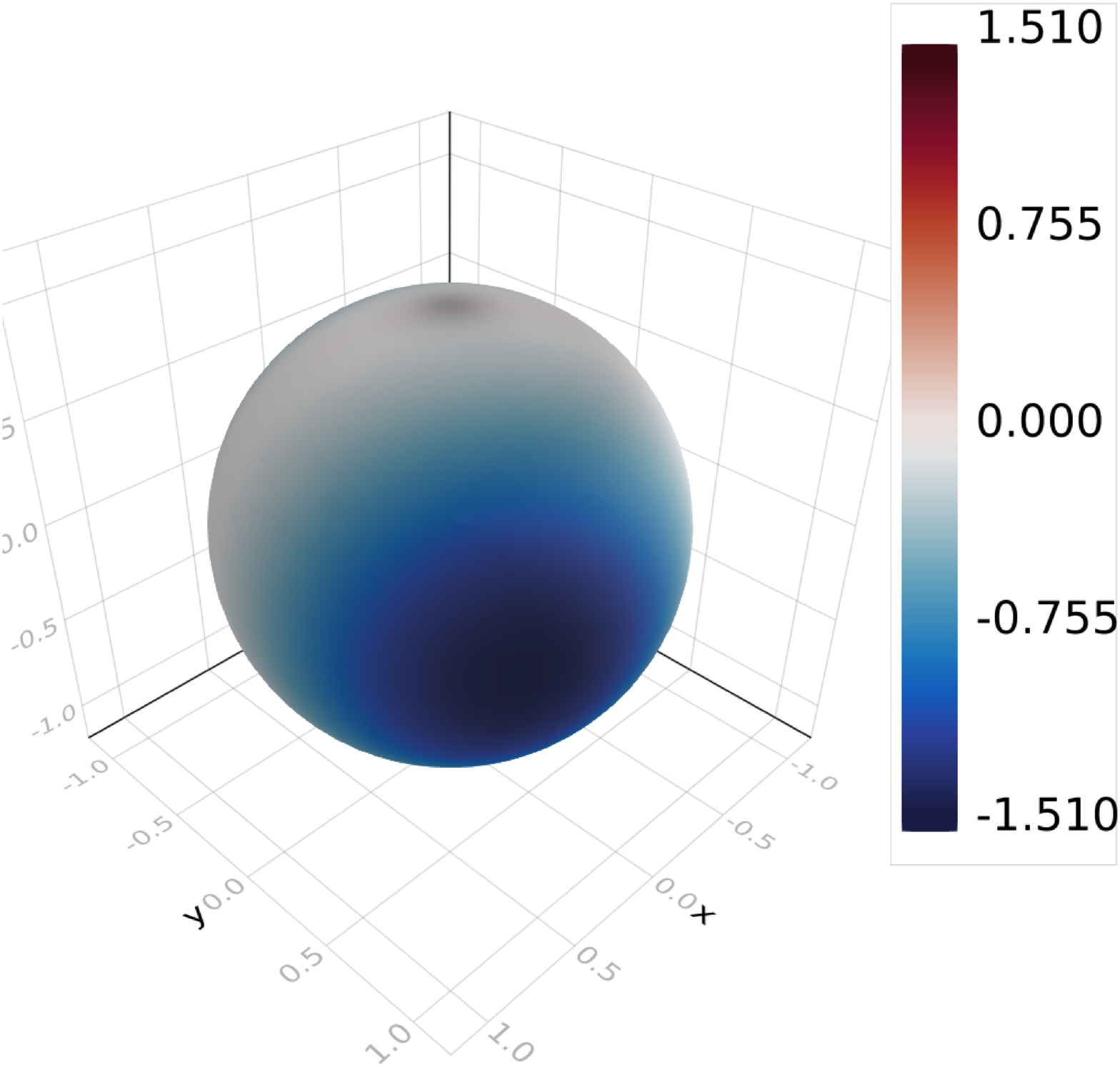}
		\put(0,85){(b)}
		\end{overpic}
		\end{center}
		\end{minipage}
	\end{tabular}
	\begin{minipage}{1.0\hsize}
	\begin{center}
	\begin{overpic}[clip,width=8.4 cm]{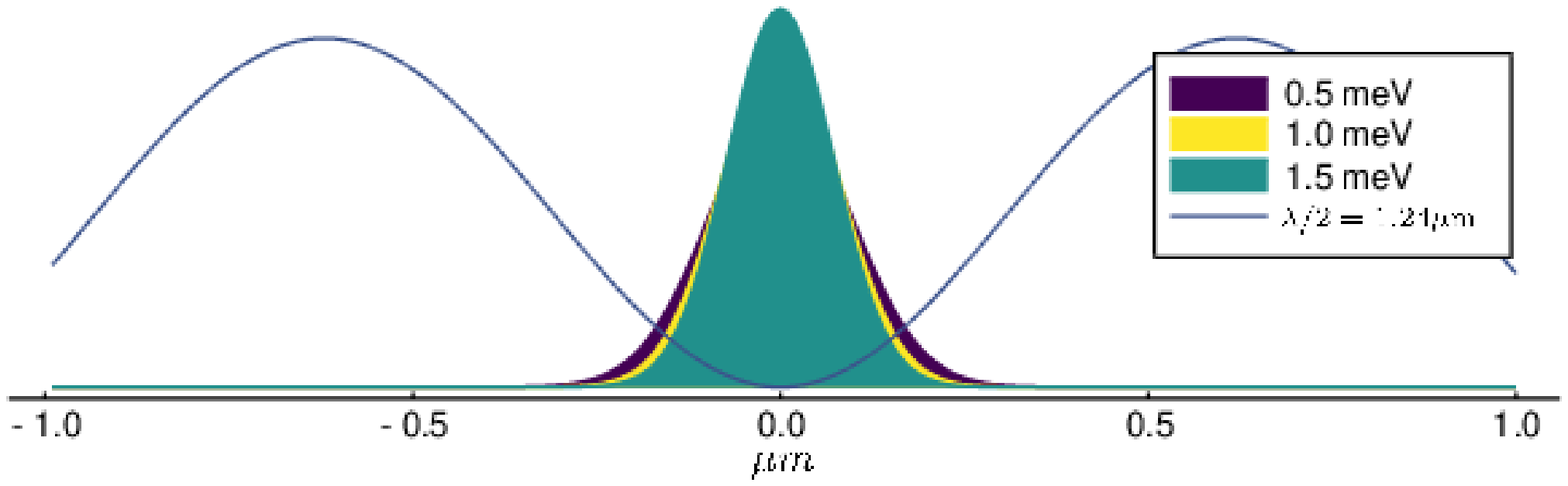}
	\put(0,30){(c)}
	\end{overpic}
	\end{center}
	\end{minipage}		
	\caption{Polarization angle dependency of the $U_{opt}^+$ for (a) two-level system and (b) three-level system in units of meV plotted on a unit sphere. The former is composed of the $E_u$ and $E_g$ excitons, and the latter is composed of the $E_u$ and $A_{2g}$ excitons. Field amplitude is $10^{-4}$ a.u. which is equal to the field intensity 3.51 $\times 10^8$ ${\rm W/cm^2} $ in (a), and it is 3.51 $\times 10^6$ ${\rm W/cm^2 } $ in (b). The detuning $\Delta$= -10 meV in both (a) and (b). $z$ axis is take to be normal to crystal plane. (c) Square of the center-of-mass wave function $|\psi_{cm}(\bm{r}_{cm})|^2\propto \exp\{-r_{cm}^2/R^2\}$ for three $U_{opt}^+$ values in the $E_u$-$E_g$ two-level system. $R$= 0.135, 0.113, 0.102 $\mu$m for  $U_{opt}^+$=0.5, 1.0, 1.5 meV, respectively. The blue line is the spatial profile of $U_{op}^+(\bm{r})$ with wave length $\lambda=1.24$ $\mu$m just for an eye guide. 
	}
	\label{fig:shift}
\end{figure}

Now we show the polarization dependency of the potential depth $U_{op}^+$ on a unit sphere in FIG.~\ref{fig:shift} for the red detuning $\Delta = -10$ meV.  
With this detuning $U_{op}^+$ is attractive whose depth reaches a few meV. 
This is comparable with the strain-induced potential in bulk crystal \cite{BHSPW07-Strain, YCK11-BECStrain}.
Even in this condition, the Rabi frequency $\Omega_R$ and detuning $\Delta$ 
are far smaller than the energy level spacing $\omega$, which validates the rotational wave approximation.
It is in virtue of following two factors : (i) large transition dipole moment and (ii) huge energy level spacing.
Particularly the factor (ii) owes to the weak screening of Coulomb potential unique in low-dimensional systems.
Another illuminating feature is the anisotropy of $U_{op}^+$ as a consequence of anisotropic dipole.
In FIG.~\ref{fig:shift}(a) $U_{op}^+$ reaches maximum depth when the polarization vector of light is oriented normal to the crystal plane. 
In this level configuration the external field does not break pseudo spin symmetry of the $E_u$ exciton.
 In the $E_u$-$A_{2g}$ three level system, the transition dipole is oriented along the crystal plane.
Significantly, the degeneracy is broken in this system; it could open up a way to manipulate the pseudo spin degrees of freedom of the $E_u$ exciton like the stimulated Raman adiabatic passage (STIRAP) \cite{KTS98,GZ14,GZ15} besides possible pseudospin selective trapping.
We have to note that the angular dependency is unitary invariant in FIG.~\ref{fig:shift} while
the $E_u$ excitonic states to span the dressed states can be varied by a unitary transformation in each  BSE calculation.

Here we consider some obstacles for trapping excitons in the optical confinement potential.
A major process that could disrupt the trapping is the radiation heating by the recoil energy $E_{rec}$ of the single photon emission process. 
Cudazzo {\it et al.} \cite{CSGRSM2010} theoretically found that the dispersion relation of the graphane exciton is well approximated by a quadratic curve with effective exciton mass $m_{exc}\sim 1.8 m_{e^-}$ where $m_{e^-}$ is the free electron mass.
$E_{rec}$ can then be of the order of $\hbar^2\bm{K}^2/2m_{exc} = 1.36\times 10^{-4} $ meV for the $E_{g} \rightarrow E_{u}$ deexcitation where $\bm{K}$ is the momentum of a photon.
The net influx of recoil energy $R_{heat}$ in unit time to the $E_u$ exciton can be approximated by a product of the inverse lifetime $\Gamma$ of deexcitation and the occupation $\rho$ of $E_g$ exciton in a dressed state as $R_{heat} \sim E_{rec}\rho\Gamma$.
We referred to the formula of the exciton lifetime by Spataru et al. on the basis of the first principles calculation \cite{CATR10}, summed up the contributions from the deexcitation processes from the $E_g$ to $E_u$ exciton states with respect to the momentum transfer $\bm{q}$, obtaining $\Gamma=5/3\cdot e^2d^2\omega^3/\hbar c^3$ and consequently its lifetime $\tau=3.5 \times 10^2$ ns.
This is much longer than the lifetime of $E_u$ exciton estimated to be 15 ps \cite{CATR10}, we thus conclude that the radiation heating effect is irrelevant since it is negligibly small in the time scale of the exciton lifetime.

Another concern is the  spatial scale of the optical confinement potential for excitons.
When you apply a standing wave of monochromic laser, it serves as a potential whose periodicity is  half a wave length, which in this case be around 1$\mu$m.% 
This scale is comparable with the typical size of  monolayer crystal flakes of two dimensional materials available experimentally.
The spatial extension of the center-of-mass  motion of the trapped exciton depends on the optical potential depth.
As a rough estimation we approximate the optical dipole trap by a 1D harmonic potential as $U_{opt}^+\cos(2\pi r/\lambda) \simeq U_{opt}^+ \{1-(2\pi r/\lambda)^2/2\}$ with the radial coordinate $r$, where $\lambda \sim 1.24$ $\mu$m for the $E_u-E_g$ system.
Then the radius $R$ of center-of-mass wave function $\psi_{\rm cm}(\bm{r}_{\rm cm})$ ( $|\psi_{cm}(\bm{r}_{cm})|^2\propto \exp\{-r_{cm}^2/R^2\}$ ) become $R=0.137$ $\mu{\rm m}$ for $|U_{op}^+|=0.5$ meV (FIG.~\ref{fig:shift}(c) ).
In the experiment of \cite{YCK11-BECStrain}, the Cu$_2$O paraexciton inflow to 
the center of the potential trap of 100 $\mu$m width is observed, where the exciton life time $\tau_{\rm Cu_2O}\sim$ 300 ns and potential gradient is the order of 0.01 meV/$\mu$m.
The graphane exciton lifetime $\tau_{\rm graphane}$ is expected to be
 $\tau_{\rm graphane} = 0.5\times10^{-4}\tau_{\rm Cu_2O}$ 
 while the dipole trap's spatial scale is 1/100 of it and potential gradient is 100 times larger.
Thus we can expect the inflow of the graphane excitons to the center of optical dipole trap. 
It will lead to the formation of a bright spot due to the deexcitation-induced light emission.

Thus far we examined the feasibility of the exciton dipole trap in analogy with the optical trapping technique of cold atom system. 
There are, however, some effects unique in condensed matter systems such as those of phonons.
The exciton-phonon interaction possibly works as either heating or cooling process depending on the system temperature, though {\it ab initio} calculations of exciton-phonon scattering ratio are expected to be numerically unaffordable.
Accordingly it is out of our scope at this moment.  

In conclusion, we theoretically explored the feasibility of a novel optical technique to manipulate the center-of-mass and pseudospin degrees of freedom of exciton in graphane. 
We firstly employed GW+BSE calculation in combination with DFT to determine the model parameters of the light-exciton interacting system.    
We found that the $E_u$, $E_g$ and $A_{2g}$ excitons form effective two- and three-level systems since they are well isolated from adjacent higher energy levels.
We also developed a theoretical method to compute the electronic transition dipole moment between excitonic states which is analogous to the Berry phase formula for ground states of periodic systems.  
Our formulation is free from the problem of phase degree of freedom of Bloch states in DFT and hence essential to obtain the dipole matrix of exciton in periodic systems.  
The dipole moments of the graphane exciton are one order larger than those of atomic systems and surely merit the implementation of optical potential.
The two-dimensional confinement of the graphane electronic states leads to anisotropic dipoles.
We clarified that the application of normal-to-plane polarized laser light equally traps the doubly degenerate $E_u$ exciton while in-plane polarized laser light induces a potential breaking of this degeneracy, {\it i.e.} pseudospin degree of freedom , when the laser frequency is blue detuned in both cases.
Such potentials can be arbitrarily shaped by tuning the spatial profile and configuration of laser setup and thus expected to advance the research of, $e.g.$ pseudospin relaxation processes or formation of exciton many-body phases.  
We also emphasize that the formulation of the transition dipole moment we introduced above will help the theoretical study of the light-exciton interacting systems driven by first principles calculations.
The exciton superradiance and/or dipole-dipole force effect are important many-body processes of excitons mediated by photon emission and reabsorption \cite{GZ14,GZ15} , where we expect that our formula is straightforwardly applicable.     

Part of this work is based on the results obtained from the NEDO project ``Development of advanced laser processing with intelligence based on high-brightness and high-efficiency laser technologies'' (TACMI project).

\bibliography{reference}
\end{document}